\documentclass{PoS}
\usepackage{setspace}
\usepackage{slashed}
\usepackage{verbatim}    
\usepackage{graphicx} 
\usepackage{wrapfig}
\usepackage{dsfont}
\usepackage{epsf}
\usepackage{epsfig}
\usepackage{epstopdf}
\usepackage{amsmath}
\usepackage{amsfonts,amssymb}
\usepackage{cleveref}
\usepackage{xcolor}

\newcommand{\be}{\begin{equation}}
\newcommand{\ee}{\end{equation}}
\newcommand{\bea}{\begin{eqnarray}} 
\newcommand{\eea}{\end{eqnarray}}
\newcommand{\MSbar}{{\overline{\rm MS}}}

\newcommand{\la}{\lambda}
\newcommand{\Dslash}{{\not{\hspace{-0.1cm}D}}}

\title{Noether supercurrent operator mixing from lattice perturbation theory}

\ShortTitle{Noether supercurrent operator mixing}

\author{G.~Bergner$\,^a$, \speaker{M.~Costa}$^{,\,b,\, c}$, H.~Panagopoulos$^{\,b}$, I.~Soler$\,^a$, G.~Spanoudes$\,^d$\\
	\llap{}
	$^a$University of Jena, Institute for Theoretical Physics,\\
     Max-Wien-Platz 1, D-07743 Jena, Germany\\
	$^b$Department of Physics, University of Cyprus,\\
1 Panepistimiou Avenue, 2109 Aglantzia, Nicosia, Cyprus \\
	$^c$Department of Chemical Engineering, Cyprus University of Technology, \\ 30 Archbishop Kyprianou Str., 3036, Limassol, Cyprus \\
	$^d$Computation-based Science and Technology Research Center,\\
The Cyprus Institute, 20 Kavafi Str., Nicosia 2121, Cyprus \\
	{\rm E-mail}:  
\email{georg.bergner@uni-jena.de},
\email{kosta.marios@ucy.ac.cy},
\email{panagopoulos.haris@ucy.ac.cy},
\email{ivan.soler.calero@uni-jena.de},
\email{g.spanoudis@cyi.ac.cy}}

\abstract{In this work we present perturbative results for the renormalization of the supercurrent operator, $S_\mu$, in ${\cal N} =1$ Supersymmetric Yang-Mills theory. At the quantum level, this operator mixes with both gauge invariant and noninvariant operators, which have the same global transformation properties. In total, there are $13$ linearly independent mixing operators of the same and lower dimensionality. We determine, via lattice perturbation theory, the first two rows of the mixing matrix, which refer to the renormalization of $S_\mu$, and of the gauge invariant mixing operator, $T_\mu$. To extract these mixing coefficients in the $\MSbar$ renormalization scheme and at one-loop order, we compute the relevant two-point and three-point Green’s functions of $S_\mu$ and $T_\mu$ in two regularizations: dimensional and lattice. On the lattice, we employ the plaquette gluonic action and for the gluinos we use the fermionic Wilson action with clover improvement.
\begin{center}
\includegraphics[scale=0.45]{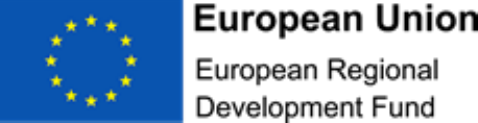}
\includegraphics[scale=0.45]{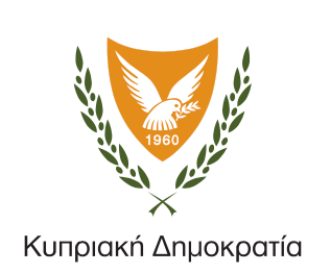}
\includegraphics[scale=0.45]{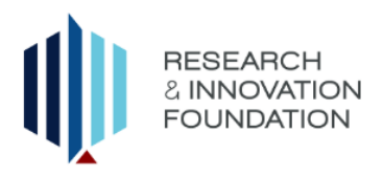}
\includegraphics[scale=0.45]{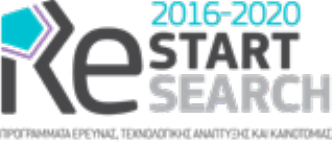}
\end{center}
}

\FullConference{%
The 39th International Symposium on Lattice Field Theory,\\
8th-13th August, 2022,\\
Rheinische Friedrich-Wilhelms-Universität Bonn, Bonn, Germany
}

\begin{document}

\maketitle

\section{Introduction to ${\cal N} = 1$ SYM and definition of operators}

The ${\cal N} = 1$ Supersymmetric Yang-Mills (SYM) Lagrangian~\cite{Curci:1986sm} connects gluon and gluino fields; it shares some of the fundamental properties of supersymmetric gauge theories containing quarks and squarks, while at the same time it is amenable to high-accuracy numerical simulations~\cite{Ali:2018fbq}. Therefore it is an ideal forerunner for future investigations of theories containing more superfields. In the Wess-Zumino (WZ) gauge,  the Lagrangian is\footnote{In order to quantize the theory, we fix the gauge by including a gauge-fixing term, together with the compensating ghost field ($c^\alpha$) terms; these terms are the same as in the non-supersymmetric case.}:
\begin{equation} 
\mathcal{L}_{\rm SYM}=-\frac{1}{4}u_{\mu \nu}^{\alpha}u_{\mu \nu}^{\alpha}+\frac{i}{2}\bar{\lambda}^{\alpha}\gamma^{\mu}\mathcal{D}_{\mu}\lambda^{\alpha}, \quad  u_{\mu \nu}= \partial_{\mu} u_\nu - \partial_{\nu} u_\mu + i g [u_\mu, u_\nu],\quad \mathcal{D}_{\mu}\lambda = \partial_{\mu}\lambda + ig [u_\mu,\lambda] 
\label{susylagr},
\end{equation}
where $u_{\mu \nu}$ is the gluon field tensor, $u_\mu$ is the gluon field, $\lambda$ is the gluino field which is a Majorana spinor in the adjoint representation of the gauge group. ${\cal L}_{\rm SYM}$ remains invariant, up to a total derivative, under the supersymmetric transformations:
\bea
\delta_\xi u_\mu^{\alpha} & = & -i \bar \xi \gamma^\mu \lambda^{\alpha}, \nonumber \\
\delta_\xi \lambda^{\alpha} & = & \frac{1}{4} u_{\mu \nu}^{\alpha} [\gamma^{\mu},\gamma^{\nu}] \xi. 
\label{susytransfDirac}
\eea
Noether's theorem gives a supercurrent for this Lagrangian stemming from the transformations of Eq.~(\ref{susytransfDirac}); in Euclidean space, the supercurrent takes the form:
\be
S_\mu = -\frac{1}{2}{\rm tr}_c ( u_{\rho\,\sigma} [\gamma_\rho,\gamma_\sigma] \gamma_\mu \lambda ) \nonumber
\ee
In this work, we make use of the Wilson formulation on the lattice, with the addition of the clover (SW) term for gluino fields. For the lattice discretization of $S_\mu$, the lattice version of gluon field tensor,  $\hat{F}_{\rho \sigma}$, which we adopt, is a sum of plaguettes in the $\rho - \sigma$ plane having $x$ as their initial and final point (see, e.g., Ref.~\cite{Bergner:2022wnb} for standard notation).

A proper study of $S_\mu$ must address the fact that it mixes with a number of other operators at the quantum level. These operators must necessarily have the same transformation properties under global symmetries (e.g. Lorentz, or hypercubic on the lattice, global $SU(N_c)$ transformations, ghost number, etc.) and their dimension must be lower than or equal to that of $S_\mu$, namely $7/2$. There are altogether four classes of such operators, as follows:
\begin{description}
\item [Class G:] Gauge-invariant operators.
\item [Class A:] BRST variation of operators.
\item [Class B:] Operators which vanish by the equations of motion.
\item [Class C:] Other operators which do not belong to the above classes.
\end{description}

$S_\mu$ being gauge invariant operator belongs to Class G. In particular,  $S_\mu$ mixes with  another dimension $7/2$ gauge invariant operator, denoted here as: 
\be
{T_\mu} = 2 {\rm tr}_c (u_{\mu\,\nu} \gamma_\nu \lambda) 
\ee
A total of twelve gauge noninvariant operators could in principle mix with $S_\mu$. These operators necessarily belong to classes A, B and C\footnote{Operators ${\cal O}_{C5}$ and ${\cal O}_{C9}$, taken together with ${\cal O}_{A1}$, are linearly dependent; however, keeping both of them in the list affords us with additional consistency checks.}:
\bea
\label{BRSToperator}
{\cal O}_{A1} &=&  \frac{1}{\alpha} {\rm tr}_c ((\partial_{\nu} u_\nu) \gamma_\mu \la ) - ig\, {\rm{tr}}_c( [c, \bar{c}]   \gamma_\mu \la) \nonumber\\[2ex] \nonumber
{\cal O}_{B1} &=& {\rm tr}_c (u_\mu \Dslash \la ),\,\,  {\cal O}_{B2} = {\rm tr}_c (\slashed{u} \gamma_\mu \Dslash \la ) \\[2ex]
{\cal O}_{C1} &=&  {\rm tr}_c (u_\mu \la) ,\,\,
{\cal O}_{C2} = {\rm tr}_c (\slashed{u}  \gamma_\mu \la),\,\,
{\cal O}_{C3} = {\rm tr}_c (\slashed{u} \partial_{\mu} \la),\,\,
{\cal O}_{C4} =  {\rm tr}_c ((\partial_{\mu} \slashed{u}) \, \la) \nonumber\\[2ex]
{\cal O}_{C5} &=& {\rm tr}_c ((\partial_{\nu} u_\nu) \gamma_{\mu} \la),\,\, 
{\cal O}_{C6} =  {\rm tr}_c (u_\nu \gamma_{\mu} \partial_{\nu} \la),\,
{\cal O}_{C7} = i\, g\ {\rm tr}_c ([u_{\rho\,},u_{\sigma}] [\gamma_\rho,\gamma_\sigma] \gamma_\mu  \la) \nonumber\\[2ex]
{\cal O}_{C8} &=& i\,g\ {\rm tr}_c ( [u_{\mu},u_{\nu}] \gamma_\nu \lambda ),\,
{\cal O}_{C9} =i\,g\ {\rm tr}_c ([c, \bar{c}] \gamma_\mu \la)
\label{All_operators}
\eea
Among them, there is just one class A acceptable operator. Two of them are not present on-shell because are class B operators. Further, there are nine class C operators, where the first two class C operators are lower dimensional operators and they only may show up in the lattice regularization. 

For a comprehensive presentation of our results, along with detailed explanations and a longer list of references, we refer to our publication~\cite{Bergner:2022wnb}.

\section{Computational setup for the renormalization of the supercurrent operator}

Our calculation set up shares a backbone of methodology, which is briefly described in three steps:
\begin{enumerate}
    \item The first step is to produce a minimal list of all candidate mixing operators by exploiting certain symmetries of the action, valid both in the continuum and the lattice formulation of the theory. This reduces the number of the operators that can possibly mix with $S_\mu$ at the quantum level to a minimum set of $13$ operators.
    \item Secondly, we careful select and compute a set of Green's functions both in the continuum and the lattice regularizations. The lattice calculations are the crux of this work; and the continuum calculations serve as a necessary introductory part, allowing us to relate our lattice results to the $\MSbar$ scheme. An unambiguous extraction of all mixing coefficients and renormalization constants of the operator $S_\mu$ entails specific choices of the external momenta for the Green's functions. In particular, we calculate two-point and three-point Green's functions of $S_\mu$ using both dimensional regularization (continuum), where we regularize the theory in $D$-dimensions ($D = 4 - 2\epsilon$), and lattice regularization. The continuum Green's functions will be used in order to calculate the renormalized Green's functions in the $\MSbar$ scheme, which are necessary ingredients for the renormalization conditions on the lattice.
    \item Lastly, we apply renormalization conditions in the $\MSbar$ scheme to the Green's functions, in order to get the results on the renormalization and the mixing coefficients. 
\end{enumerate}  
Note that the same mixing operators may mix with $T_\mu$. We follow the same methodology and we also calculate the same one-loop Green's functions with insertion of the $T_\mu$ operator. Thus, we determine the second row of the $13 \times 13$ mixing matrix. However these results~\cite{Bergner:2022wnb} are omitted here for the sake of brevity.

For off-shell matrix elements, the mixing assumes this general form:
\begin{equation}
S_\mu^R  = Z_{SS} S_\mu^B +Z_{ST} T_\mu^B + Z_{SA1} {\cal O}_{A1}^B + \sum_{i=1}^{2} Z_{SBi} {\cal O}_{Bi}^B  + \sum_{i=1}^{9} Z_{SCi} {\cal O}_{Ci}^B,
\label{Zz} 
\end{equation}
where the superscript $B$ stands for the bare and $R$ for renormalized quantities. In order to determine all Z-factors, we consider two-point Green's functions with one external gluino and one external gluon fields ($\langle u_\nu^{\alpha_1}({-q_1})\,S_\mu({x})\, \bar\lambda^{\alpha_2}({q_2} ) \rangle$), as well as three-point Green's functions with external gluino/gluon/gluon fields ($\langle u_\nu^{\alpha_1}({-q_1})\, u_\rho^{\alpha_2}({-q_2} ) \,S_\mu({x})\, \bar\lambda^{\alpha_3}({q_3}) 
	    \rangle$) and with gluino/antighost/ghost fields ($\langle c^{\alpha_3}({q_3}) \,S_\mu({x})\,\bar c^{\alpha_2}({q_2} )  \bar\lambda^{\alpha_1}({-q_1}) \rangle$); similarly for $T_\mu$.

In Table~\ref{tb:treelevels}, we show the tree-level Green's functions for all operators apart from overall color and exponential factors. These functions naturally show up in the results of the bare Green’s functions of $S^R_\mu$, allowing us to deduce the corresponding mixing coefficients. Furthermore, the tree-level Green's functions with the same external fields may depend on more than one external momentum $q_i$; this is a consequence of the absence of momentum conservation since there is no summation/integration over the position of the operators. Although this seems to complicate things it is a way to disentangle the mixing patterns. For this reason, it is convenient to calculate the Green's functions for specific choices of the external momenta. Taking into account potential IR divergences, which may appear at exceptional values of $q_i$, a sufficient set of choices for external momenta are: 3 choices for the two-point Green's functions with external $u(q_1) \la(q_2)$: $q_2 = 0$, $q_1= 0$, ${q_2} = - {q_1}$, as well as a single choice for each of the two three-point Green's functions with external $u(q_1) u(q_2) \la( {q_3})$ and $\la({q_1}) \bar c ({q_2}) c({q_3}) $: (${q_2} = 0$, ${q_3} = - q_1$) and ($q_2 = q_1$, $q_3 = 0$), respectively.

\begin{table}[ht!]
\begin{center}
 \begin{tabular}{c | c | c | c} 
 \hline
 & Tree Level two-point  & Tree Level three-point & Tree Level three-point  \\
 Operators & Green's function& Green's function& Green's function \\
 & (external legs: $u_\nu({-q_1})\,\bar \la({q_2})$) & (external legs: $u_\nu\,u_\rho \, \bar \la$)& (external legs: $c\,\bar c \, \bar \la$) \\ [0.5ex] 
 \hline\hline
$S_\mu$ & $-i (\slashed{q_1} \gamma_\nu - q_{1\nu})\gamma_\mu $  & $\,g\,[\gamma_\nu,\gamma_\rho] \gamma_\mu/2$ & $0$\\ [2ex]
 \hline
$T_\mu$ & $ i(q_{1\mu} \gamma_\nu - \slashed q_{1} \delta_{\mu \nu}  )$  & $ - \,g \, (\delta_{\mu \nu} \gamma_\rho + \delta_{\mu \rho} \gamma_\nu) $ & $0$\\ [2ex]
 \hline
 ${\cal O}_{A1}$ & $i\,q_{1\nu}\gamma_\mu/(2\alpha)$ & $0$ & $(g/2) \gamma_\mu$ \\ [2ex]
 \hline
 ${\cal O}_{B1}$ & $i\,\delta_{\mu \nu} \slashed{q_2}/2 $ & $-\,g\,(\delta_{\nu \mu} \gamma_\rho + \delta_{\rho \mu} \gamma_\nu)/2$ & $0$ \\[2ex]
 \hline
 ${\cal O}_{B2}$ & $i\,\gamma_\nu \gamma_\mu \slashed{q_2}/2 $ &  $-2\,g\,\gamma_\nu \gamma_\mu \gamma_\rho$ & $0$ \\[2ex]
 \hline
 ${\cal O}_{C1}$ & $\delta_{\mu \nu}/2$ & $0$ & $0$ \\[2ex]
 \hline
 ${\cal O}_{C2}$ & $\gamma_\nu \gamma_\mu/2$ & $0$ & $0$ \\ [2ex] 
\hline
 ${\cal O}_{C3}$ & $i\,\gamma_\nu q_{2\mu}/2$ & $0$ & $0$ \\[2ex]
 \hline
 ${\cal O}_{C4}$ & $i\,\gamma_\nu q_{1_\mu}/2$ & $0$ & $0$ \\ [2ex] 
\hline
 ${\cal O}_{C5}$ & $i\,\gamma_\mu q_{1\nu}/2$ & $0$ & $0$ \\[2ex]
 \hline
 ${\cal O}_{C6}$ & $i\,\gamma_\mu q_{2\nu}/2$ & $0$ & $0$ \\ [2ex] 
\hline
 ${\cal O}_{C7}$ & $0$ & $-g\,[\gamma_\nu,\gamma_\rho] \gamma_\mu$ & $0$ \\[2ex]
 \hline
 ${\cal O}_{C8}$ & $0$ & $-\,g\,(\delta_{\nu \mu} \gamma_\rho + \delta_{\rho \mu} \gamma_\nu)/2$ & $0$ \\ [2ex] 
\hline
 ${\cal O}_{C9}$ & $0$ & $0$ & $-(g/2) \gamma_\mu$ \\[2ex]
  \hline \hline
\end{tabular}
\caption{Two-point and three-point amputated tree-level Green's functions of $S_\mu$ and $T_\mu$, as well as of gauge noninvariant operators which may mix with $S_\mu$\label{tb:treelevels}. $\langle u_\nu^{\alpha_1}(-q_1)\,{\cal O}_{i}(x)\, \bar\lambda^{\alpha_2}(q_2) \rangle$, $\langle u_\nu^{\alpha_1}(-q_1)\, u_\rho^{\alpha_2}(-q_2)\,{\cal O}_{i}(x)\, \bar\lambda^{\alpha_3}(q_3) \rangle$ and $\langle c^{\alpha_3}(q_3) \,{\cal O}_{i}(x)\,\bar c^{\alpha_2}(q_2) \bar\lambda^{\alpha_1}(q_1) \rangle$ are shown apart from overall factors of $\delta^{\alpha_{1}\alpha_{2}} e^{i\,x\cdot(q_1+q_2)}$, $f^{\alpha_{1}\alpha_{2}\alpha_{3}} e^{i\,x\cdot(q_1+q_2+q_3)}$ and $f^{\alpha_{1}\alpha_{2}\alpha_{3}} e^{i\,x\cdot(q_1-q_2+q_3)}$, respectively.}
\end{center}
\end{table}

\begin{figure}[ht!]
\centering
\includegraphics[scale=0.3]{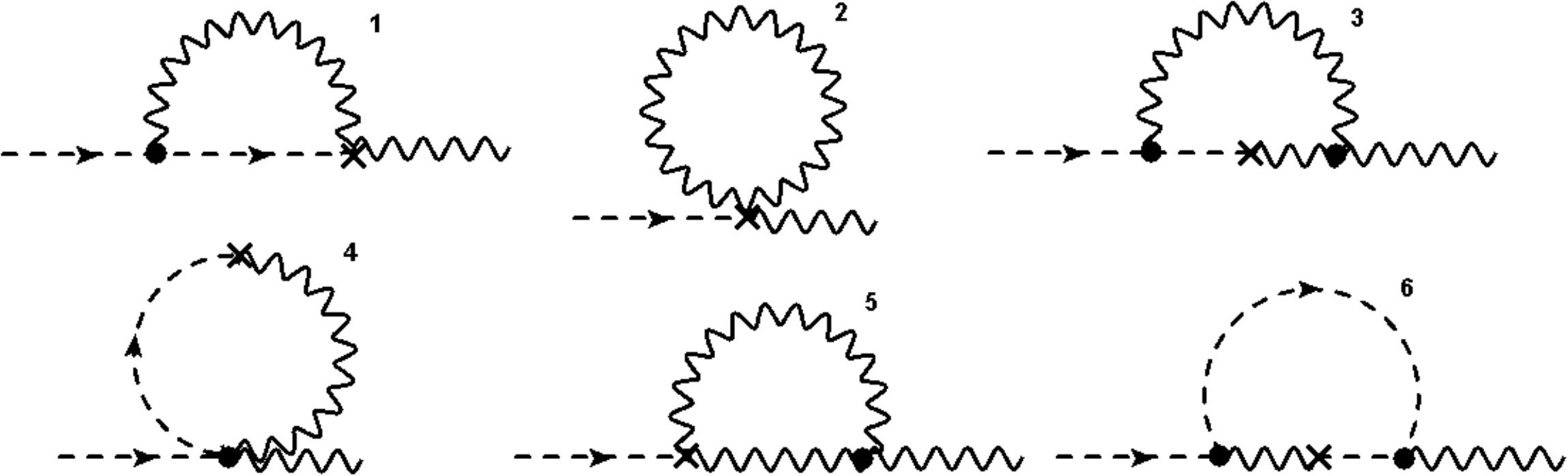}
\caption{One-loop Feynman diagrams contributing to the two-point Green's functions $\langle u_\nu S_\mu \bar \la  \rangle$\, and $\langle u_\nu T_\mu \bar \la  \rangle$.  A wavy (dashed) line represents gluons (gluinos). A cross denotes the insertion of $S_\mu$($T_\mu$). Diagrams 2, 4 do not appear in dimensional regularization; they do however show up in the lattice formulation.}
\label{fig2pt}
\end{figure}

\begin{figure}[ht!]
\centering
{\includegraphics[scale=0.3]{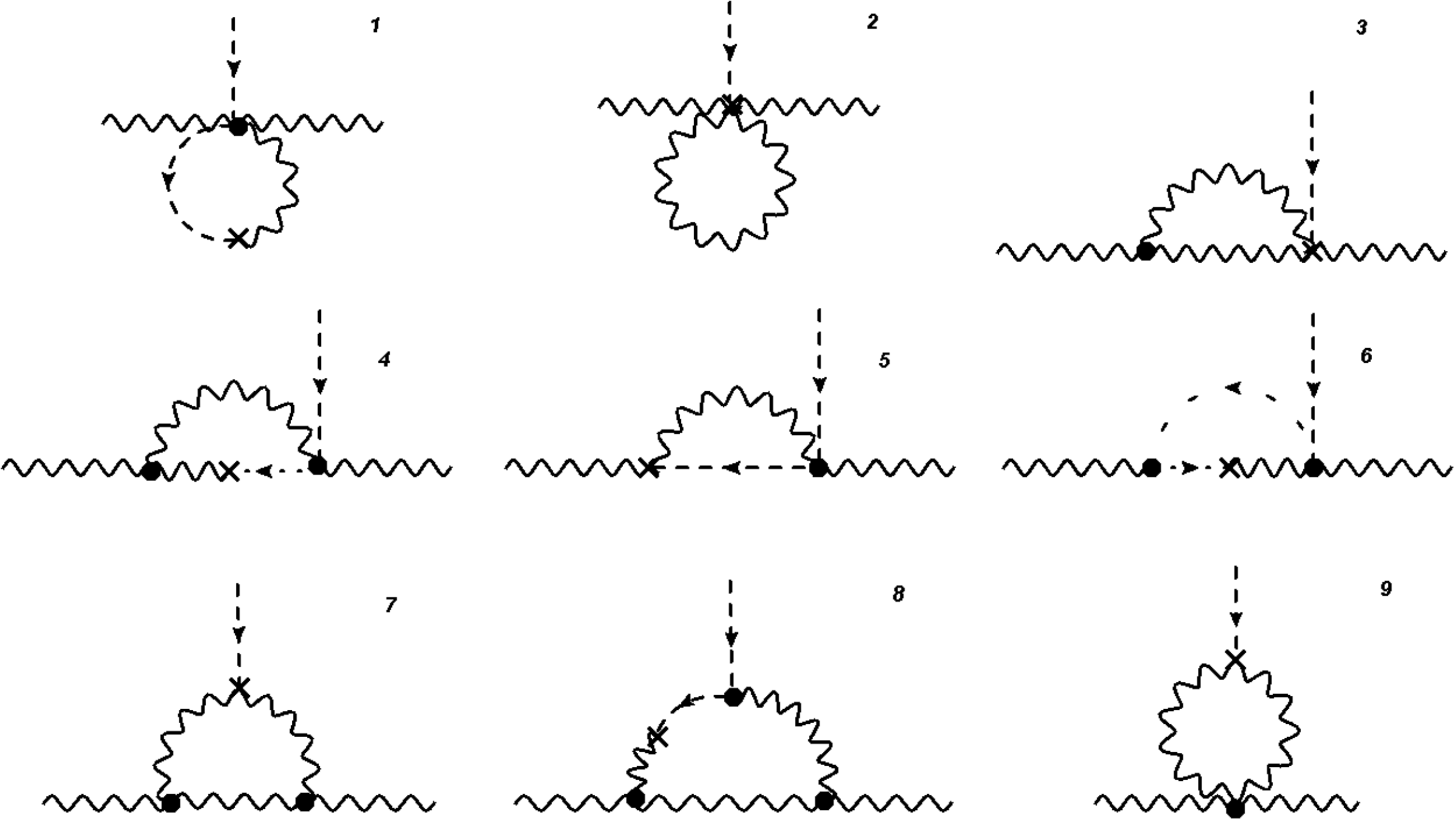}\\
\includegraphics[scale=0.3]{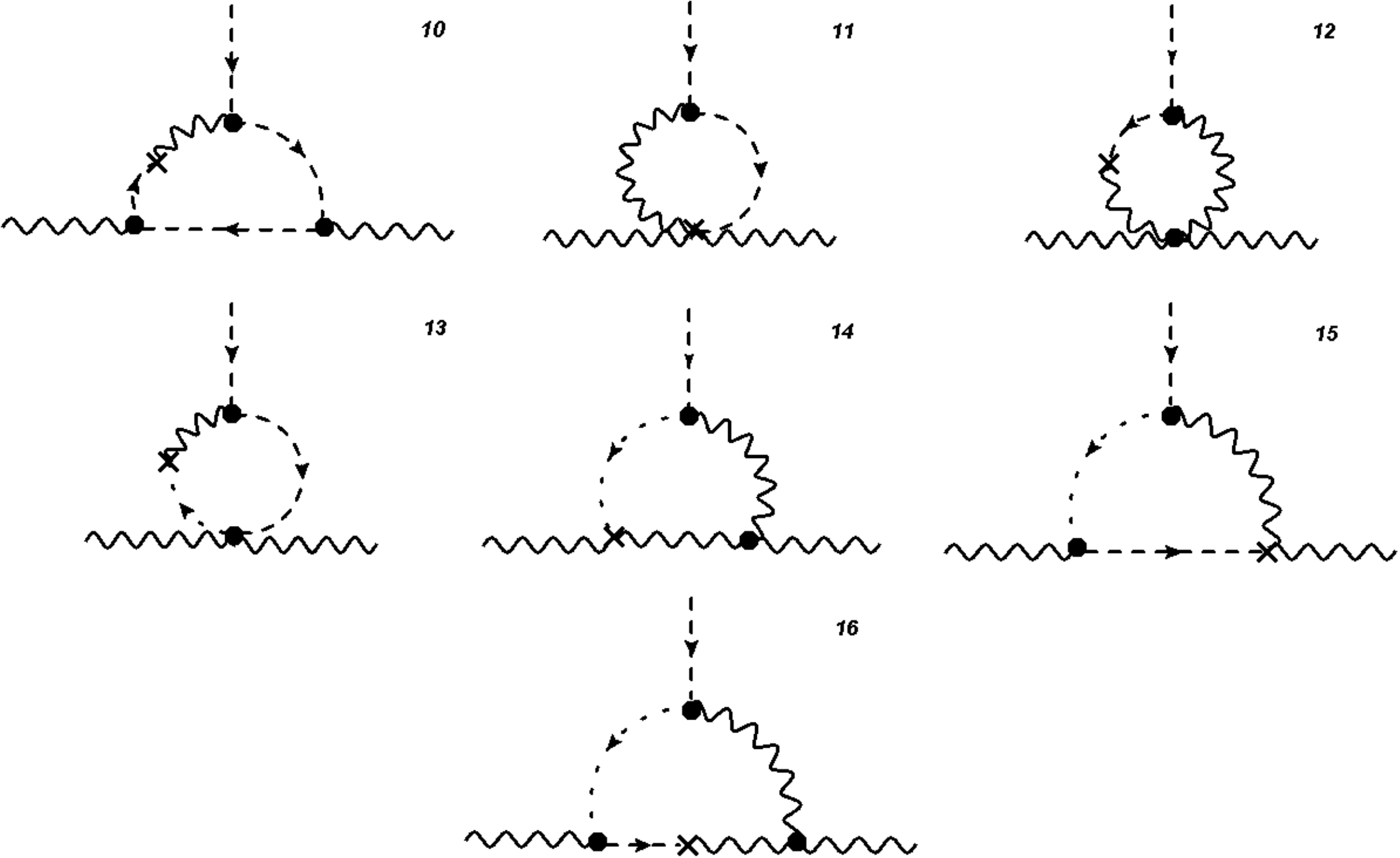}}
\caption{One-loop Feynman diagrams contributing to the three-point Green's functions $\langle u_\nu u_\rho S_\mu\bar \la  \rangle$\, and $\langle u_\nu u_\rho T_\mu\bar \la  \rangle$\,.  Diagrams 1, 2, 3, 5, 6, 11, and 13 do not appear in dimensional regularization but they contribute in the lattice regularization. A mirror version  of diagrams 3, 4, 5, 6, 8, 10, 14, 15 and 16 must also be included.}
\label{fig3ptguuF}
\end{figure}

\begin{figure}[ht!]
\centering
\includegraphics[scale=0.3]{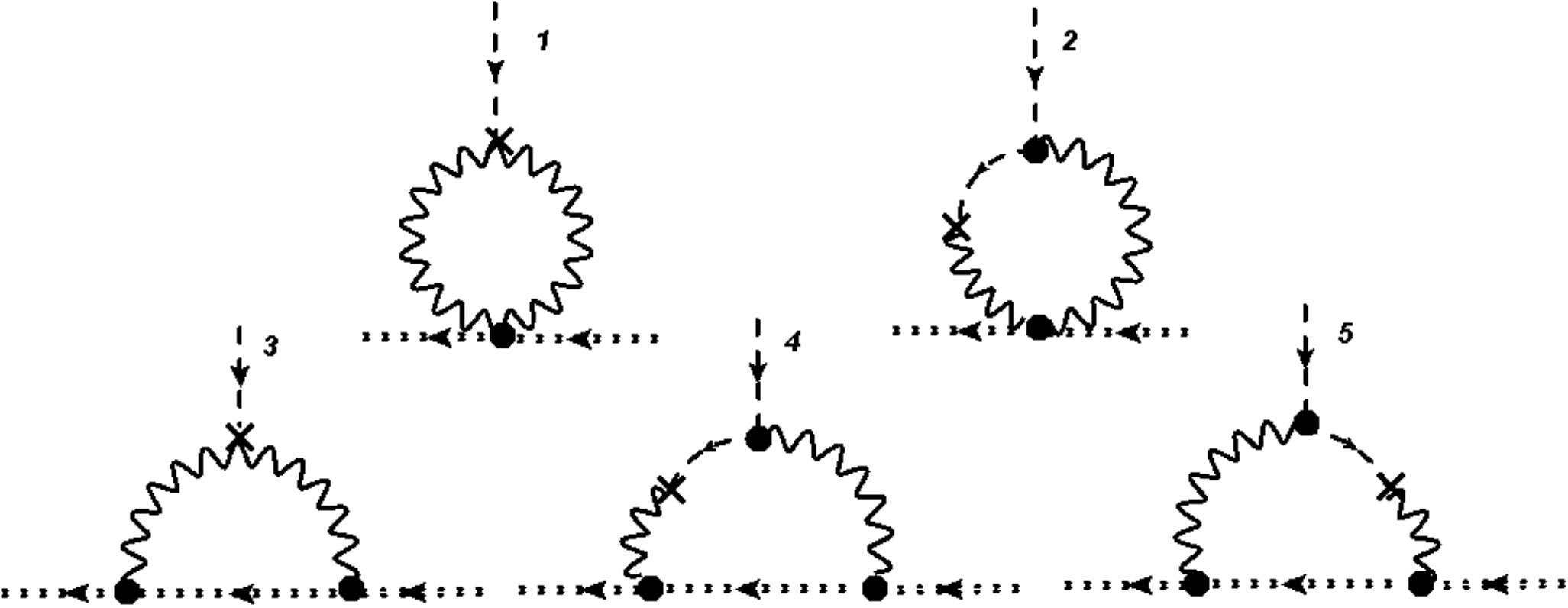}
\caption{One-loop Feynman diagrams contributing to the three-point Green's functions $\langle c \, S_\mu \, \bar c \, \bar \la \rangle$\, and $\langle c \, T_\mu \, \bar c \, \bar \la \rangle$\,.  The ``double dashed'' line is the ghost field. Diagrams 1 and 2 do not appear in dimensional regularization; they do however show up in the lattice formulation.}
\label{fig3ptgCC}
\end{figure}

Each Green’s function is written as a sum of several Feynman diagrams. The one-loop Feynman diagrams (one-particle irreducible (1PI)) contributing to corresponding Green’s functions are shown in Figures~\ref{fig2pt}, \ref{fig3ptguuF}, \ref{fig3ptgCC}.

\section{Results for Green's functions and for the mixing matrix on the lattice}

Both $\MSbar$-renormalized and bare Green's functions have the same tensorial structures.  At the one-loop order the differences between the $\MSbar$-renormalized and corresponding bare lattice Green's functions appear in the renormalization conditions in the $\MSbar$ scheme. These differences are polynomial in the external momenta and proportional to the tree-level Green's functions of the operators. Due to the fact that these differences appear in the renormalization conditions, we present them here. First of all, the resulting expression for the difference between the two-point $\MSbar$-renormalized and lattice bare Green's functions of $S_\mu$ for $q_2=0$, is:

\bea
\langle u_\nu^{\alpha_1}(-q_1)  \,{S}_{\mu}\, \bar\lambda^{\alpha_2}(q_2)  \big|^{\MSbar}_{q_2 = 0}  - \langle u_\nu^{\alpha_1}(-q_1)  \,{S}_{\mu}\, \bar\lambda^{\alpha_2}(q_2)  \rangle  \big|^{LR}_{q_2 = 0} &=& i\frac{g^2}{16\pi^2}\frac{1}{2} \delta^{\alpha_1\,\alpha_2} e^{iq_{1}x} N_c \times \Bigg[ \nonumber\\ 
&& \hspace{-9cm}-5.99999\, \slashed q_{1}  \delta_{\mu\,\nu} +\gamma_{\nu} q_{1\mu} 5.99722 + (\gamma_{\nu} \gamma_{\mu} \slashed q_1 + \gamma_\mu q_{1\nu} - 2 \gamma_{\nu} q_{1\mu} )\bigg(\frac{39.47842}{N_c^2} - 30.57429  \nonumber \\ 
&& \hspace{-9cm}+ 5.17830\alpha-  4.55519 c_{\rm SW}^2 + 5.3771 c_{\rm SW} r + \frac{3}{2} (1- \alpha) \log\left(a^2 \bar \mu^2\right)\bigg) \Bigg]
\eea
The absence of $q$-independent terms means that the lower-dimensional operators, ${\cal O}_{C1}$ and ${\cal O}_{C2}$, do not mix with $S_\mu$. The renormalization condition for the two-point Green's functions involves the renormalization factors of the external fields and of the supercurrent operator as well as the corresponding nonvanishing tree-level Green's functions along with their mixing coefficients. The condition applied to the gluino-gluon Green's function of the operator ${S}_{\mu}$ reads to one loop:


\bea 
\langle u_\nu^R \,{S}^R_{\mu}\, \bar\lambda^R  \rangle   &=& Z_\la^{-1/2} \,Z_u^{-1/2} \langle u_\nu^B \,{S}^R_{\mu}\, \bar\lambda^B \rangle \nonumber\\ 
& =& Z_\la^{-1/2} \,Z_u^{-1/2} Z_{SS} \langle u_\nu^B \,{S}^B_{\mu}\,\bar\lambda^B  \rangle  + Z_{ST}  \langle u_\nu^B \, {T_\mu}^B\, \bar\lambda^B  \rangle^{tree} \nonumber\\ 
&+& Z_{SA1} \langle  u_\nu^B \, {\cal O}_{A1}^B \, \bar\lambda^B  \rangle  ^{tree}\nonumber\\ 
&+& \sum_{i=1}^{2} Z_{SBi} \langle  u_\nu^B \, {\cal O}_{Bi}^B \, \bar\lambda^B  \rangle  ^{tree} + \sum_{i=1}^{6} Z_{SCi} \langle u_\nu^B \,{\cal O}_{Ci}^B \, \bar\lambda^B  \rangle  ^{tree} + {\cal O}(g^4) 
\label{2ptGFexprS}
\eea
 
From the choice $q_2 = 0$ we extract:
\bea
Z_{SS}^{LR,\MSbar} &=& 1 + \frac{g^2}{16\pi^2}(\frac{-9.86960}{N_c} + N_c(-2.3170 + 14.49751c_{\rm SW}^2 - 1.23662c_{\rm SW}\,r))\\
Z_{ST}^{LR,\MSbar} &=& \frac{g^2}{16\pi^2} 3N_c\\
Z_{SA1}^{LR,\MSbar}  &=&  Z_{SC1}^{LR,\MSbar} = Z_{SC2}^{LR,\MSbar}=Z_{SC4}^{LR,\MSbar} = Z_{SC5}^{LR,\MSbar}  = 0 
\eea
An important feature of the supercurrent operator is that its renormalization is finite: this is in line with its classical conservation. The mixing with $T_\mu$ on the lattice is related to the $\gamma$-trace anomaly of the supercurrent operator and it is in agreement with older results in the literature~\cite{Taniguchi:1999fc}. Further, there is no mixing with ${\cal O}_{A1}$, ${\cal O}_{C4}$ and ${\cal O}_{C5}$ operators.

Since for the choice $q_2=0$ the tree-level two-point Green's functions of ${\cal O}_{B1}, {\cal O}_{B2}, {\cal O}_{C3}, {\cal O}_{C6}$ vanish, we evaluate the two-point Green's functions at $q_1=0$, leading to this expression:
\bea
\label{firstdiff}
\langle u_\nu^{\alpha_1}(-q_1)  \,{S}_{\mu}\, \bar\lambda^{\alpha_2}(q_2)  \rangle  \big|^{\MSbar}_{q_1 = 0}  - \langle u_\nu^{\alpha_1}(-q_1)  \,{S}_{\mu}\, \bar\lambda^{\alpha_2}(q_2)  \rangle  \big|^{LR}_{q_1 = 0} &=&  i\frac{g^2 N_{c}}{16\pi^2}\times \nonumber \\ 
&& \hspace{-11cm} \frac{1}{2} \delta^{\alpha_1\,\alpha_2}e^{iq_2x}\times \Bigg[
\gamma_{\nu} \gamma_{\mu} \slashed q_2 \bigg(0.80802   - \frac{1}{2} \log\left(a^2 \bar \mu^2\right) \bigg) - \slashed q_{2}  \delta_{\mu\,\nu} \bigg(0.38395+ \log\left(a^2 \bar \mu^2\right)\bigg)  \Bigg]
\eea

From this choice ($q_1 = 0$) we determine the logarithmically divergent mixings with class B operators as well as with ${\cal O}_{C3}$ and ${\cal O}_{C6}$.
\bea
Z_{SB1}^{LR,\MSbar} &=& \frac{g^2}{16\pi^2} N_c \left(-0.38395 - \log\left(a^2 \bar \mu^2\right)\right)\label{ZSB1}\\
Z_{SB2}^{LR,\MSbar} &=&  \frac{g^2}{16\pi^2} N_c \left(0.80802 -\frac{1}{2} \log\left(a^2 \bar \mu^2\right)\right)\label{ZSB2}\\
Z_{SC3}^{LR,\MSbar} &=& Z_{SC6}^{LR,\MSbar}  = 0
\eea

All the previous results are consistent with the choice $q_2=-q_1$. The expression for the corresponding difference at $q_2 = - q_1$ is:
\bea
\langle u_\nu^{\alpha_1}(-q_1)  \,{S}_{\mu}\, \bar\lambda^{\alpha_2}(q_2)  \rangle  \big|^{\MSbar}_{q_2 = -q_1}  - \langle u_\nu^{\alpha_1}(-q_1)  \,{S}_{\mu}\, \bar\lambda^{\alpha_2}(q_2)  \rangle  \big|^{LR}_{q_2 = -q_1} &=& i\frac{g^2}{16\pi^2}\frac{1}{2} \delta^{\alpha_1\,\alpha_2}  N_c \times \Bigg[
\nonumber \\
&& \hspace{-10cm}
 0.80802 \gamma_\mu q_{1\nu} + 4.38396\gamma_\nu q_{1\mu}
+\frac{1}{2} \gamma_{\nu} \gamma_{\mu} \slashed q_1\log\left(a^2 \bar \mu^2\right)
+ (\gamma_{\nu} \gamma_{\mu} \slashed q_1 + \gamma_\mu q_{1\nu}- 2\gamma_\nu q_{1\mu} ) \bigg(\frac{39.47842}{N_c^2}\nonumber \\ && \hspace{-9cm} -31.38231 + 5.17830 \alpha 
- 4.55519 c_{\rm SW}^2 + 5.37708 c_{\rm SW} r  + 2 \log\left(a^2 \bar \mu^2\right) -  \frac{3}{2} \alpha \log\left(a^2 \bar \mu^2\right) \bigg)\nonumber \\
&&\hspace{-5cm}+\slashed q_{1} \delta_{\mu\,\nu} \left(-5.61605 +\log\left(a^2 \bar \mu^2\right) \right)
\Bigg]
\label{lastdiff}
\eea

In order to determine the mixing coefficients with the operators ${\cal O}_{C7}$, ${\cal O}_{C8}$ and ${\cal O}_{C9}$, we also need to impose a set of renormalization conditions on three-point Green's functions.  The first one involves two external gluons and one gluino:
\bea
\langle u_\nu^R u_\rho^R \,{S}^R_{\mu}\, \bar\lambda^R  \rangle   &=&Z_\la^{-1/2} \,Z_u Z_{SS} \langle u_\nu^B u_\rho^B \,{S}^B_{\mu}\,\bar\lambda^B  \rangle   + Z_{ST} \langle u_\nu^B u_\rho^B \, T_\mu^B\, \bar\lambda^B  \rangle  ^{tree} \nonumber\\
&+&  \sum_{i=1}^{2} Z_{SBi}\langle u_\nu^B u_\rho^B \, {\cal O}_{Bi}^B\, \bar\lambda^B  \rangle  ^{tree} +\sum_{i=7}^{8} Z_{SCi}\langle u_\nu^B u_\rho^B \, {\cal O}_{Ci}^B\, \bar\lambda^B  \rangle  ^{tree} + {\cal O}(g^4)
\label{3ptGFexpr1S}
\eea
The second one involves external gluino/antighost/ghost fields:
\bea 
\langle c^R \,{S}^R_\mu\,\bar c^R\, \bar\lambda^R  \rangle   &=& Z_c^{-1} Z_\la^{-1/2} Z_{SS} \langle c^B {S}^R_\mu\bar c^b \bar\lambda^B   \rangle   + Z_{ST} \langle c^B  T_\mu^B\bar c^B \bar\lambda^B  \rangle  ^{tree} \nonumber\\
&+&Z_{SA1}\langle c^B \, {\cal O}_{A1}^B\,\bar c^B \bar\lambda^B  \rangle  ^{tree} +  Z_{SC9}\langle c^B \, {\cal O}_{C9}^B\,\bar c^B \bar\lambda^B  \rangle  ^{tree} + {\cal O}(g^4)
\label{3ptGFexpr2S}
\eea

The result for the difference of the first three-point Green's function is:
\bea
&&\hspace{-1.5cm}\langle u_\nu^{\alpha_1}(-q_1) u_\rho^{\alpha_2}(-q_2) \,{S}_{\mu}\, \bar\lambda^{\alpha_3}(q_3) \rangle \big|^{\MSbar}_{q_2 =0 , q_3 = -q_1} -\langle u_\nu^{\alpha_1}(-q_1) u_\rho^{\alpha_2}(-q_2) \,{S}_{\mu}\, \bar\lambda^{\alpha_3}(q_3) \rangle \big|^{LR}_{q_2 =0 , q_3 = -q_1} =\nonumber \\
&& \hspace{0cm}
 \frac{g^3 N_c}{16\pi^2}\,f^{\alpha_{1}\alpha_{2}\alpha_{3}}\, \Bigg[
\left(\delta_{\nu \rho} \gamma_\mu-\gamma_\nu \gamma_\rho \gamma_\mu \right)\bigg(\frac{19.73920}{N_c^2} - 12.48660 + 3.28231 \alpha  \nonumber \\
&&\hspace{3cm}
- 2.27761 c_{\rm SW}^2 + 2.68854 c_{\rm SW} r 
+ \frac{1-2\alpha}{2} \log \left(a^2 \bar \mu^2\right)\bigg) 
\Bigg]
\label{3ptGGgLattS}
\eea
With this result we conclude that there are no mixings with ${\cal O}_{C7}$ ad ${\cal O}_{C8}$. The lattice Green's functions containing gluino-ghost-antighost external fields are identical to the continuum ones at one loop order; thus, there is also no mixing of $S_\mu$ with ${\cal O}_{C9}$.

\section{Summary -- Conclusion}

To summarize the main points, we have seen that the supercurrent operator suffers from mixing with both gauge invariant and noninvariant operators. We have  calculated the first two rows of the mixing matrix; these correspond to the two gauge invariant operators which are involved. The results are important in order to have a thorough picture of the mixing pattern when gauge noninvariant and off-shell Green's functions are employed. The novelty in our one-loop results is that we calculate the complete mixing patterns of the supercurrent operator perturbatively. More precisely, we use gauge-variant off-shell Green's functions; we obtain analytic expressions for the renormalization factors and mixing coefficients, where the number of colors $N_c$, the coupling constant $g$, the gauge parameter $\alpha$, the clover/Wilson parameters $c_{\rm SW}/ r$ (on the lattice) are left unspecified.

In our ongoing investigations, the renormalization of $S_\mu$ is also deduced by calculating exclusively gauge-invariant Green's functions; in this case the mixing of gauge-noninvariant operators is irrelevant and there is no need to fix the gauge or introduce ghost fields, leaving only the effective $2 \times 2$ space of $S_\mu$ and $T_\mu$ mixings. Therefore, this Gauge-Invariant Scheme (GIRS) is more accessible via non-perturbative calculations. For detailed information on non-perturbative results, see the proceedings by I. Soler in this conference \cite{Ivan}.\\\\
{\bf Acknowledgements:} 
 G.B. and I.S. acknowledge financial support from the Deutsche Forschungsgemeinschaft (DFG) Grant No.~432299911 and 431842497. 
 This work was co-funded by the European Regional Development Fund and the Republic of Cyprus through the Research and Innovation Foundation (Project: EXCELLENCE/0918/0066 and EXCELLENCE/0421/0025). M.C. also acknowledges partial support from the Cyprus University of Technology under the ``POST-DOCTORAL" programme. G.S. acknowledges financial support from H2020 project PRACE-6IP (Grant agreement ID: 823767).

\end{document}